\documentclass{pasa}%

\usepackage{graphicx}
\usepackage[dvipsnames]{xcolor}

\title[High Cadence Optical Transient Searches]{High Cadence Optical Transient Searches using Drift Scan Imaging I: Proof of Concept with a Pre-Prototype System}

\author[Steven Tingay]{Steven Tingay
\affil{International Centre for Radio Astronomy Research, Curtin University, Bentley, WA 6102, Australia}%
}%

\jid{PASA}
\doi{10.1017/pas.\the\year.xxx}
\jyear{\the\year}

\usepackage{aas_macros}
\usepackage{hyperref} 
\hypersetup{colorlinks,citecolor=blue,linkcolor=blue,urlcolor=blue}

\begin{document}

\begin{frontmatter}
\maketitle

\begin{abstract}
An imaging technique with sensitivity to short duration optical transients is described.  The technique is based on the use of wide-field cameras operating in a drift scanning mode, whereby persistent objects produce trails on the sensor and short duration transients occupy localised groups of pixels.  A benefit of the technique is that sensitivity to short duration signals is not accompanied by massive data rates, because the exposure time $>>$ transient duration.  The technique is demonstrated using a pre-prototype system composed of readily available and inexpensive commercial components, coupled with common coding environments, commercially available software, and free web-based services.  The performance of the technique and the pre-prototype system is explored, including aspects of photometric and astrometric calibration, detection sensitivity, characterisation of candidate transients, and the differentiation of astronomical signals from non-astronomical signals (primarily glints from satellites in Earth orbit and cosmic ray hits on sensor pixels).  Test observations were made using the pre-prototype system, achieving sensitivity to transients with 21 ms duration, resulting in the detection of five candidate transients.  An investigation of these candidates concludes they are most likely due to cosmic ray hits on the sensor and/or satellites.  The sensitivity obtained with the pre-prototype system is such that, under some models for the optical emission from FRBs, the detection of a typical FRB, such as FRB181228, to a distance of approximately 100 Mpc is plausible.  Several options for improving the system/technique in the future are described.
\end{abstract}

\end{frontmatter}

\section{INTRODUCTION }
\label{sec:introduction}
One of the exciting new areas of astrophysics to emerge over the last decade has been the study of Fast Radio Bursts (FRBs), millisecond bursts of radio emission from unknown progenitors within galaxies at cosmological distances.  The rate of discovery of new FRBs is currently very high; a database of all published FRB detections can be found at http://frbcat.org \citep{2016PASA...33...45P}.  The observed radio emission promises insights into the physics of the progenitors, probes of the conditions of the magneto-ionised material along billions of lightyears of path between the progenitors and the Earth, and possibly utility as probes of cosmology.

FRBs appear likely to be a wonderful tool for astrophysics and cosmology, however we only have knowledge of their radio emission.  What about the multi-wavelength properties of FRBs?  Searches for transient emission associated with FRBs have been conducted in various forms.  A brief review of general multi-wavelength searches for emission from FRBs has been produced by \citet{2018NatAs...2..845B}.  This paper adopts a focus on optical emission associated with FRBs, and short duration optical transients in general. 

From a theoretical modeling point of view, \citet{yang} predict the optical emission from the observed FRB radio emission, considering two different versions of Inverse Compton emission models plus an extension of the radio emission mechanism into the optical regime.  They find that under special conditions, optical emission from FRBs may be detectable under the Inverse Compton models.  Overall, \citet{yang} find the detection probablility for optical emission from FRBs to be low, but that if optical emission from even a small number of FRBs was detected it would provide valuable insights into the physics of the progenitors.  

The best chance for the optical detection of an FRB may come via the closest events.  In this sense, the recent localisation of an FRB to a spiral galaxy at 149$\pm$0.9 Mpc (luminosity distance) is encouraging \citep{2020Natur.577..190M}.  Also recently, the first evidence for a periodicity in a repeating FRB was found, an approximate 16 day period in FRB 180916.J0158$+$65 \citep{2020arXiv200110275T}.  While this period is many orders of magnitude longer than for pulsars, pulsed optical emission has been detected from a handful of pulsars \citep{2018IAUS..337..191S}.  If the periodic FRB emission is due to a pulsar-like mechanism (an option explored in the discovery paper), this also raises the prospect of detectable optical emission from some FRBs.

In this context, recent experiments have shown that the FRB event rate is high, with facilities such as CHIME \citep{2014SPIE.9145E..22B} now detecting hundreds of new FRBs.  The recently estimated event rate is 3600 FRBs/sky/day at $>$0.63 Jy at 350 MHz and 5 ms burst width \citep{2017ApJ...844..140C}.  Outside these observational parameters, many more FRBs may exist (e.g. at lower flux densities).  Thus, this event rate translates into at least 0.003 FRBs per sq. deg. per hour.  For a 10 sq. deg. detector, one FRB should occur within the field of view for every 30 hours of observation time.  This represents a substantial amount of observing time for a professional, multi-use optical telescope, and suggests that dedicated experiments are required as a first step, probably with relatively small aperture facilities, in order to accumulate time on sky and push detection limits lower.

Assuming that any transient optical emission from FRBs will also have a millisecond timescale, observations with a sub-second cadence are likely to be most useful.  Given the nature of FRBs, occuring at random times and positions on the sky, a wide field of view is also a highly desirable characteristic for optical surveys, as is integrated time on sky.  This forms a challenging parameter space to work in, implying very high data rates and data volumes from such a sensor.  Few experiments have entered this parameter space.

\citet{2019arXiv191011343R} reports on observations from the prototype Tomo-e Gozen camera with a 0.5 s cadence, over 1.9 sq. deg. (1.2$''$ pixels) and with a 105 cm diameter aperture and 59 hours of exposure time over eight nights in a variety of conditions.  No detections of astrophysical transients were made, with a limiting V magnitude of 15.6 in a 1 s exposure.  However, detections of satellites and other non-astrophysical objects were made.  \citet{2019arXiv191011343R} also lists prior similar experiments, notably the Mini-MegaTORTORA system \citep{2017ASPC..510..526K}, which achieves 0.128 s cadence, 900 sq. deg. (16$''$ pixels) with a 85 mm diameter aperture and a limiting V magnitude of 11 in a 0.128 s exposure.  Again, Mini-MegaTORTORA only detected non-astrophysical transient signals.

Aside from these experiments that may be sensitive to optical emission from FRBs but represent blind searches, targeted searches have also been possible.  In the case of FRB121102, the first known repeating FRB, \citet{2017MNRAS.472.2800H} observed the field at 70.7 ms cadence and a limiting optical flux density of 0.33 mJy, but did not detect any optical transient.  On the other hand, \citet{Tingay_2019} utilised archived TESS images that coincided in time and position with FRB181228, a non-repeating FRB, to obtain an upper limit on optical emission from the FRB of 2000 Jy for a 1 ms duration (the TESS 30 min imaging cadence, 21'' pixels, and cosmic ray mitigation procedures complicate the analysis).

While not designed for sub-second cadence, dedicated facilities for the optical detection and/or follow-up of multi-wavelength and multi-messenger transients, including FRBs and gravitational wave events, are close to operational, in MeerLICHT \citep{2019IAUS..339..203P}, BlackGEM \citep{2019lsof.confE..33G}, and Deeper, Wider, Faster \citep{2020MNRAS.491.5852A}.

Thus, limited experiments have been performed in a relevant part of the optical transient parameter space.  In this paper, an efficient technique for achieving high cadence optical imaging for short timescale transient detection is described.  Whereas the experiments listed above use short exposure times to achieve high cadence, the work described here utilises long exposures with a camera that remains pointed at a fixed azimuth and elevation, meaning that the sky drifts across the detector.  In this scenario, persistent sources of emission form trails across the detector.  However, transient sources with a duration less than $T\sim\frac{1.3751\times10^{4}p}{F{\rm cos}\delta}$ s will appear as a point source, where $p$ is the sensor pixel width (in metres), $F$ is the focal length of the camera (in metres), and $\delta$ is the declination at which the pointing direction is fixed.  Thus, with a wide field sensor and a camera of modest focal length, sensitivity to sub-second transients can be achieved using long exposure times (seconds to tens of seconds).  The downside of such a system is in localisation capability; for a sensor in which trails are formed in pixel rows (east-west), localisation in declination is good (corresponding to the width of the camera point spread function) but localisation in right ascension is worse (corresponding to twice the angular extent of the sky drift across the sensor).  

An extension to this scenario that is explored in this paper is where the camera does not remain fixed, but is actively driven at a non-siderial rate; in particular, if the pointing direction is driven in an anti-siderial direction, the higher drift rate produces higher cadence.  For example, in this case, the equation above is modified to be $T\sim\frac{1.3751\times10^{4}p}{F({\rm cos} \delta + d)}$ s, where $d$ is the angular rate (arcseconds per second) at which the pointing direction is driven in an anti-siderial sense.

Subsequent sections in this paper cover: a description of a pre-prototype system configured as a proof of concept demonstration in \S 2; a description of observations and data processing, including photometry and astrometry analyses, and a sensitivity analysis in \S3; results and discussion in \S 4; and conclusions and future work, including future planned systems and experiments that build upon the pre-prototype in \S 5.

\section{PRE-PROTOTYPE SYSTEM HARDWARE}
\label{sec:hardware}
The pre-prototype system described here is based purely on commercial-off-the-shelf (COTS) components and commonly available software tools, as a minimal system to demonstrate the approach, gain experience with data processing techniques, and obtain some initial results.

The system is built around a Canon 600D digital camera attached to a Samyang 500 mm focal length, f/6.3 lens (aperture diameter of approximately 79 mm - comparable diameter to the Mini-MegaTORTORA system described in the introduction).  The Canon 600D has a sensor size of 5190$\times$3461 pixels and a pixel size of 4.29$\mu$m, giving an approximate 2.5$^{\circ}\times$1.7$^{\circ}$ field of view.

The camera and lens were mounted on a refurbished ``heritage'' equatorial mount (Tele Vue Systems Mount RSM-2000) equipped to drive in right ascension in both siderial and anti-siderial senses at either the siderial rate or a ``fast'' rate.  The long dimension of the sensor was oriented in the north-south direction on the sky.

Test observations were conducted from a suburban Perth location: 32$^{\circ}$3'9'' S; 115$^{\circ}$54'12'' E; elevation$=$20 m) on 2019 November 16 between 15:06 and 15:16 UT.  The equatorial mount was levelled and aligned in right ascension and declination, the camera was focussed using a bright star, then the camera was pointed toward the equator and the local meridian.  A short exposure was acquired and the image was solved for astrometry by upload to http://www.astrometry.net \citep{2010AJ....139.1782L}, returning an image scale of 1.77''/pixel, a field of view of $1.7^{\circ}\times2.6^{\circ}$, an image centre of (RA,Dec)=(02h 28m 41.703s, $+$02d 05m 46.916s) at the time of the exposure, and a 1.4$^{\circ}$ offset of the sensor vertical east of north. 

\section{Observations and Data Processing}
\label{sec:dataProcessing}
A series of 11$\times$5 s exposures were made, captured from the Canon 600D to a laptop using the standard Canon software designed to control the camera over a USB connection.  During the exposures, the mount was driven in an anti-siderial direction at the ``fast'' rate (below we estimate the drive rate from the data).  Images were captured in RAW format, at approximately 19 MB per image.  Accompanying the images, 10$\times$5 s dark frames (without shutter open) were acquired in order to characterise hot pixels in the sensor.

\begin{figure*}[h!]
\begin{center}
\includegraphics[width=\textwidth]{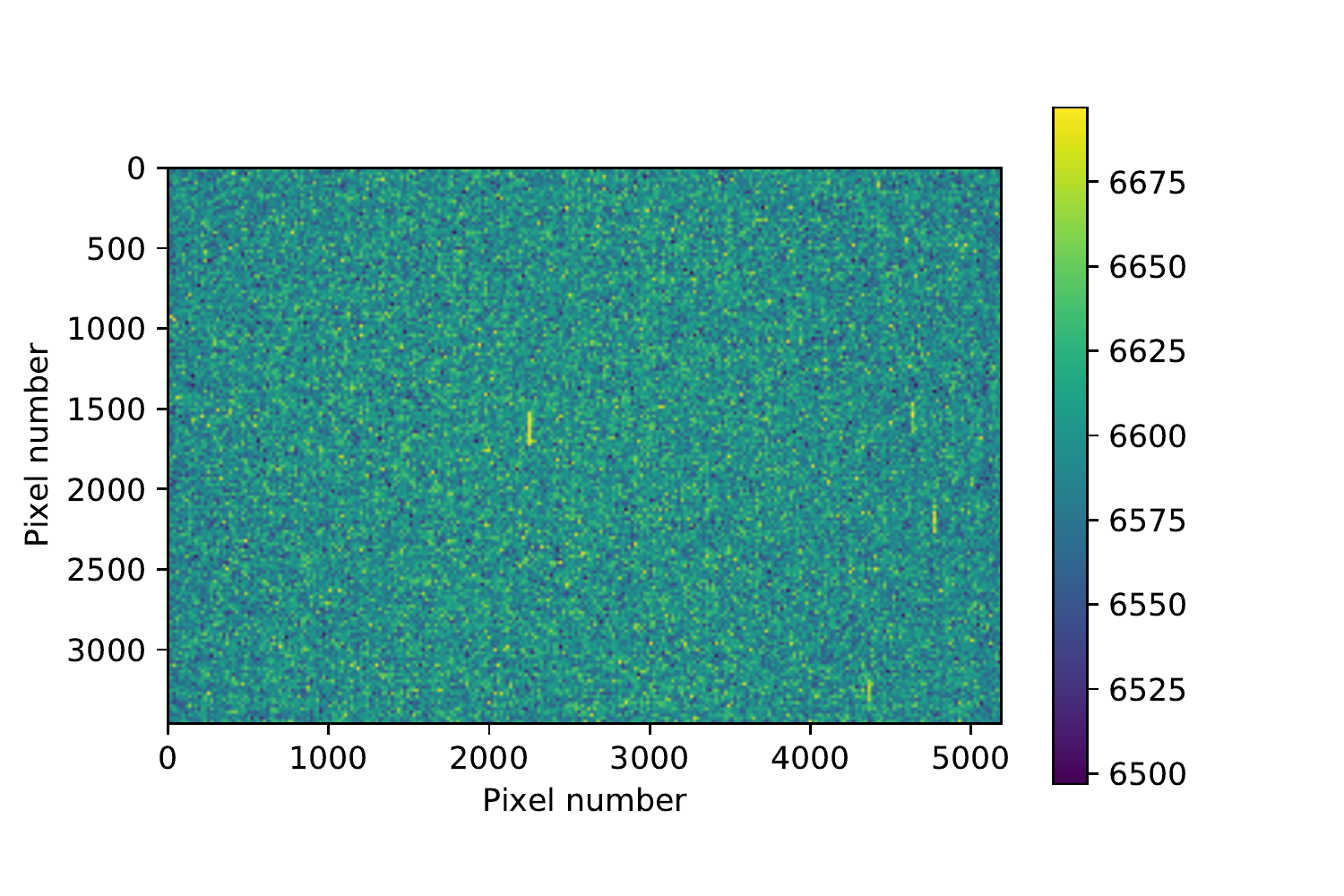}
\caption{An example 5 s image after initial data processing described in the text.  Trails from the brightest stars in the field are apparent.}
\label{image}
\end{center}
\end{figure*}

The dark frame images were processed using the Nebulosity software\footnote{https://www.stark-labs.com/nebulosity.html} to produce a ``bad pixel map'', which was then applied to each of the science images.  Using Nebulosity, the RAW science images were de-mosaiced, pixels were squared, the colour background offset was adjusted, vertical smoothing via de-interlacing of the RAW images was performed, the frames were converted to monochrome, and the images saved as FITS images.  An example image is shown in Figure \ref{image}.

\subsection{Drift rate, localisation, and timing}
From the trails produced by the brightest stars in Figure \ref{image}, we measure a trail length of 201 pixels, or 355''.  Thus, during the 5 s exposure, the drift rate of the sky across the sensor caused by driving the camera in an anti-siderial direction is 71''/s or 40 pixel/second.  Thus, the parameter $d$ in the second equation in the introduction has a value of 3.7 and $T=21$ ms; transient signals less than 21 ms in duration will appear as a point source in the images.

Ignoring the small 1.4$^{\circ}$ misalignment of the image with the north-south direction, the trails occur in an east-west direction.  Thus, if a short duration transient is detected, the localisation of the signal in the declination direction is approximately 4 pixels ($\sim$7'', estimated by forming a profile across a bright star trail and measuring its FWHM).  However, the signal could have occurred at any point during the 5 s exposure and therefore anywhere within twice the measured trail length of 710'' in right ascension. 

The timing of exposures recorded by the system, and listed in Table \ref{detections}, are only accurate to approximately 1 second, due to the time taken between exposures to save the frame to disk.

\subsection{Astrometry}
The astrometry described above was derived for a short single exposure at a time of 02:48:45 UT.  For a fixed pointing direction, the astrometry at any other time can be derived in a straightforward manner by advancing the right ascension according to the time difference between the image of interest and the reference time of 02:48:45, since the declination and sensor orientation are constant.  

However, in the case used here, the camera is driven in an anti-siderial sense at a rate that is only approximately known.  Thus, while the image centre declination and image orientation are fixed, the right ascension should be derived from the identification of stars in each image.

This was achieved by measuring the midpoints of the star trails in an unbiased manner by using a matched filter of 200 pixels length, orientated in an east-west direction.  The filter returns the mean pixel values across those 200 pixels, centred on each pixel in the image (excepting an east-west guard band of 100 pixels at each edge of the image).  The maximum values returned for each trail provide the (x,y) image coordinates for the midpoints of the detected star trails.  A threshold for detection was set at the 1$\sigma$ noise level in the image.  For example, for frame \#106 (refer Table \ref{detections}), this method returned the (x,y) coordinates for the midpoints of 10 star trails across that frame.  

These (x,y) coordinates for each frame of interest were uploaded to http://www.astrometry.net \citep{2010AJ....139.1782L} and a plate solution was obtained, returned in the form of a FITS file containing the WCS (World Coordinate System) description.  This WCS FITS file was used to derive the right ascension and declination for pixels of interest in the frames of interest (pixels containing candidate detections, see Table \ref{detections}).

\subsection{Photometry}
The sensor in the Canon 600D has a spectral sensitivity which is maximised within a wavelength range of approximately 450 nm to 600 nm, which is most closely matched to the V band out of the set of Johnston-Cousins bands.  Thus, an approximate photometric calibration of the images is possible, to determine the sensitivity performance of the system.  A set of 17 stars were detected as trails in the 11 images and the peak pixel values were measured using profiles from cuts through these trails.  The RMS near these profiles were also measured, away from the signal from the star.  The V magnitudes of the stars were obtained using the SIMBAD astronomical database \citep{2000A&AS..143....9W}, sourced from the Tycho-2 catalogue of the 2.5 million brightest stars \citep{2000A&A...355L..27H}.

\begin{figure}[h!]
\begin{center}
\includegraphics[width=0.5\textwidth]{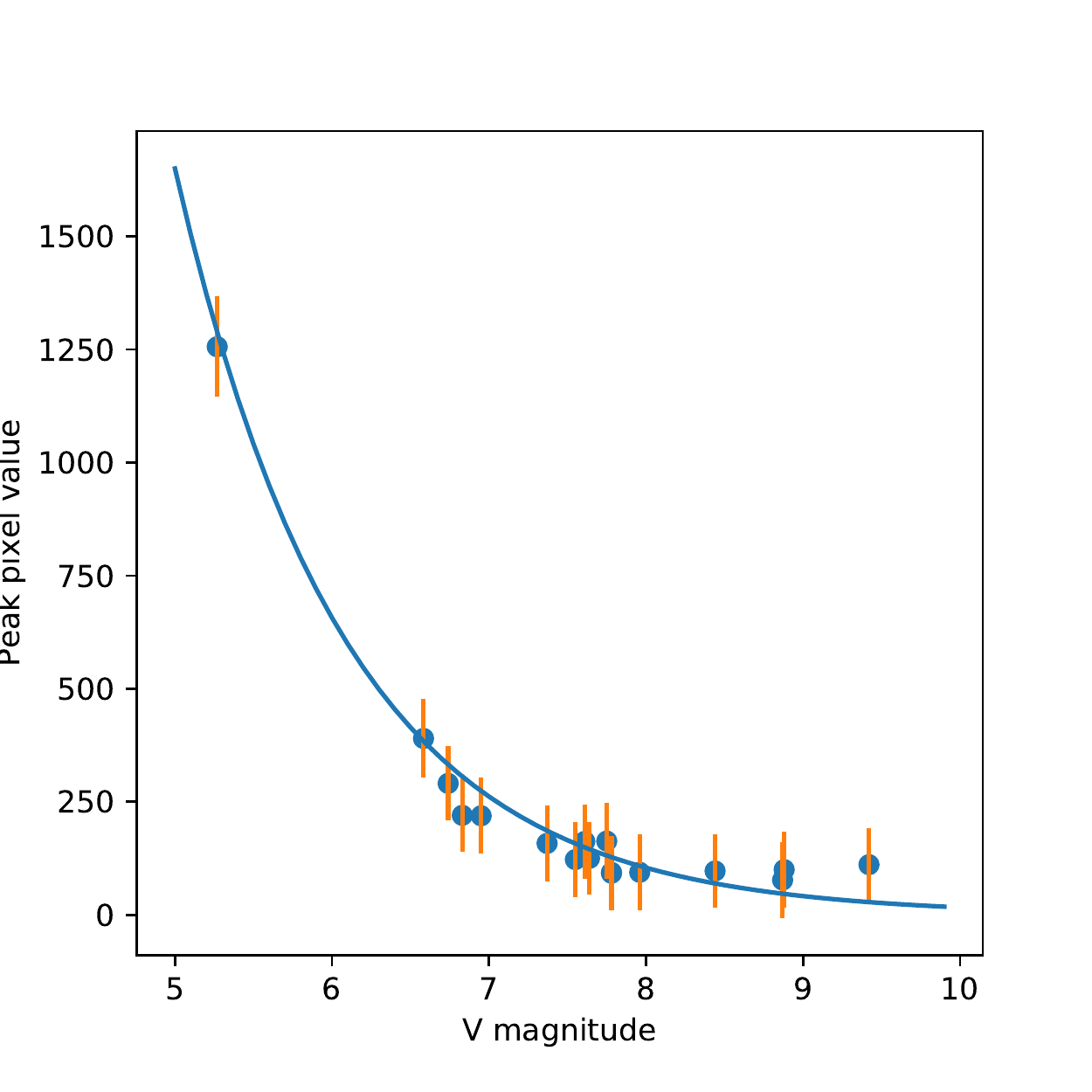}
\caption{Relationship between V magnitude and peak pixel value for 17 stars detected in the images, calibrating the photometry for these observations.  The errors shown are $\pm$3 times the measured RMS.}
\label{phot}
\end{center}
\end{figure}

Figure \ref{phot} shows the peak pixel value as a function of V magnitude, along with a function of the form $c=1.65\times10^{(5-V/2.5)}$, where $c$ is the peak pixel value and $V$ is V magnitude.  Figure \ref{phot} shows that if a transient signal is detected, the V magnitude can be measured via this calibration curve.

\subsection{Detection of candidates}
\begin{figure*}[h!]
\begin{center}
\includegraphics[width=\textwidth]{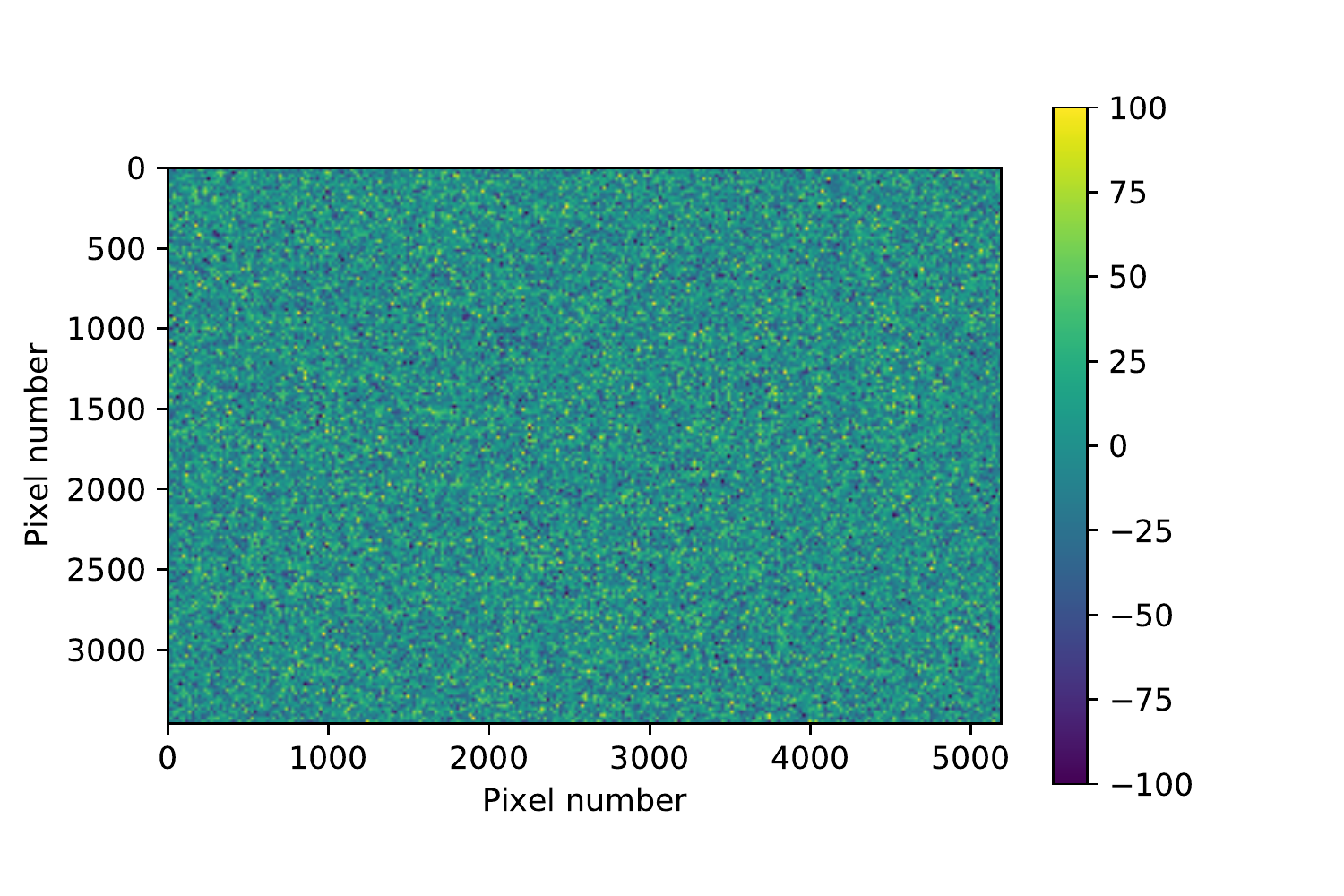}
\caption{An example difference image, formed as described in the text, from the data shown in Figure \ref{image}.}
\label{diff}
\end{center}
\end{figure*}

In order to identify candidate transient objects, point sources should be detected in the images.  In order to emphasise point source emission, remove the presence of persistent objects, remove the dark current, and flatten the noise across the images, an operation to form difference images was performed.  Pixel to pixel differences were calculated by subtracting image row $n$ from image row $n+1$, resulting in images such as seen in Figure \ref{diff}.

In such a difference image, a transient signal of duration $<$21 ms will appear as a positive point source, immediately adjacent in an east-west direction to a negative point source of approximately the same magnitude.  This is a distinct signature that allows transients to be distinguished from other types of signals.

Residual bad pixels (constantly at high values) not removed by the bad pixel maps, will have the same signature as a true transient, but will be readily identifiable due to being located in the same pixel in repeated images.

Objects such as meteors or satellites with motions on the sky will also be readily identifiable, since in general they will not move in an east-west direction and will appear in difference images as trails.  Geostationary satellites will appear as point sources with a transient signature, but like bad pixels will be detected repeatedly at the same location in multiple images.  In the case of the actively driven system, geostationary satellites will produce trails, but of shorter length than stars. Atmospheric muons caused by cosmic rays impacting the atmosphere will appear similar to hot pixels but will not repeat at the same pixel locations.  As such, signals with a cosmic ray origin require differentiation from astronomical transients.  Later in \S4, cosmic ray signals are found to be probably the dominant source of candidate signals in the data.  This fact sets considerations that will influence the future development of the prototype system in \S5.

Thermal noise fluctuations in the difference images will appear as a single positive or negative pixel and it would be exceptionally rare to find positive and negative fluctuations mimicking a transient signature in the difference image.

A simple threshold can be applied to the difference images and signals that match the signature of an expected astronomical transient can be readily identified and investigated.  Thus, the overall observation and data processing steps are relatively simple and can be efficiently implemented.  For the work described here, all steps after the initial processing in Nebulosity are implemented using standard Python packages.

\section{RESULTS AND DISCUSSION}
\begin{figure*}[h!]
\begin{center}
\includegraphics[width=\textwidth]{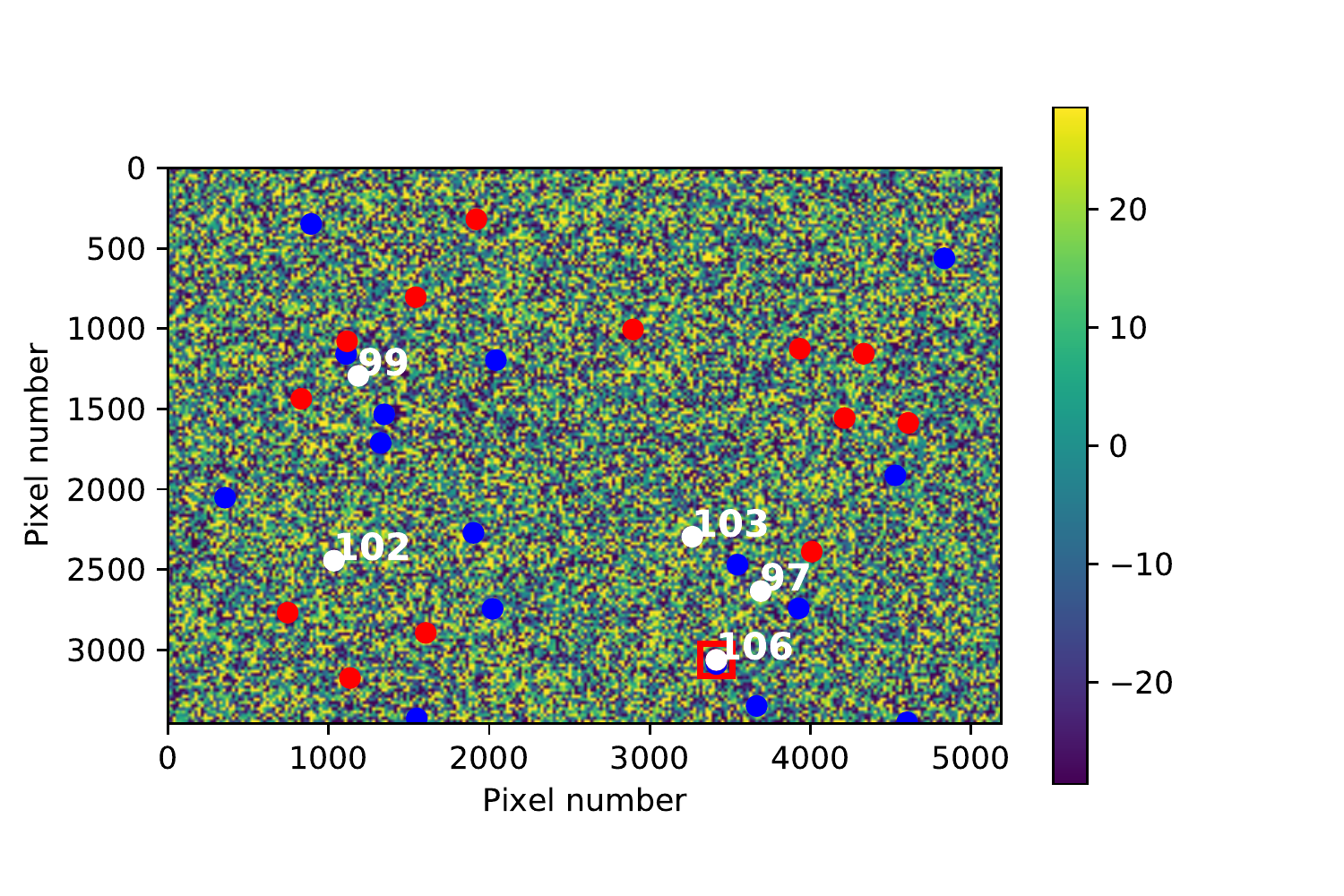}
\caption{An example difference image, overlaid with pixels $>$10 times the RMS (blue) and $<$-10 times the RMS (red).  The image is annotated with the positions of the five detections obtained over the 11 images that have the expected signature of a transient.  The numbers attached to the annotation represent the RAW frame number of the observation sequence.  The box features the detection made for this frame (\#106), which is zoomed into in Figure \ref{zoom}}
\label{diff-sig}
\end{center}
\end{figure*}

The set of 11 images was processed as described.  A threshold of 10 times the RMS was set to identify candidate objects in the difference images.  A total of five candidate transient objects were thus detected across the 11 images (no more than one in any single image) that adhered to the expected signature for a transient.  Figure \ref{diff-sig} shows a difference image, overlaid with the locations of the pixels that have counts outside the $\pm$10 times the RMS threshold (blue for positive, red for negative) for that particular image.  Figure \ref{diff-sig} also contains the locations of the five candidate signals that adhere to the signature of a transient event across the 11 images.  The five detections occur at different pixel locations, astronomical coordinates, and times.  The candidates do not appear to be residual bad pixels, as the pixel locations of the candidates were checked in all other frames and no comparable signals were found at those pixel locations.

\begin{figure}[h!]
\begin{center}
\includegraphics[width=0.5\textwidth]{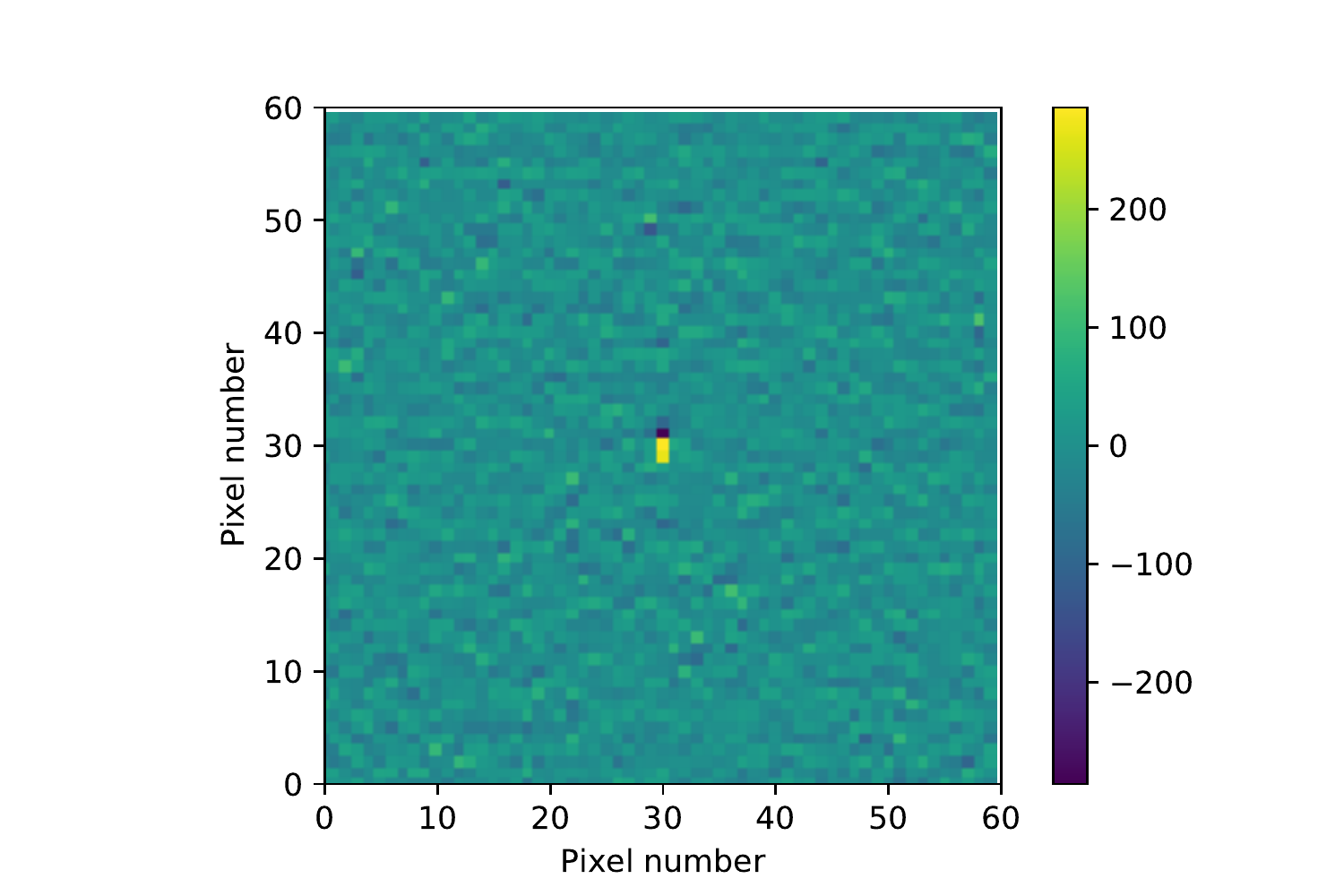}
\includegraphics[width=0.5\textwidth]{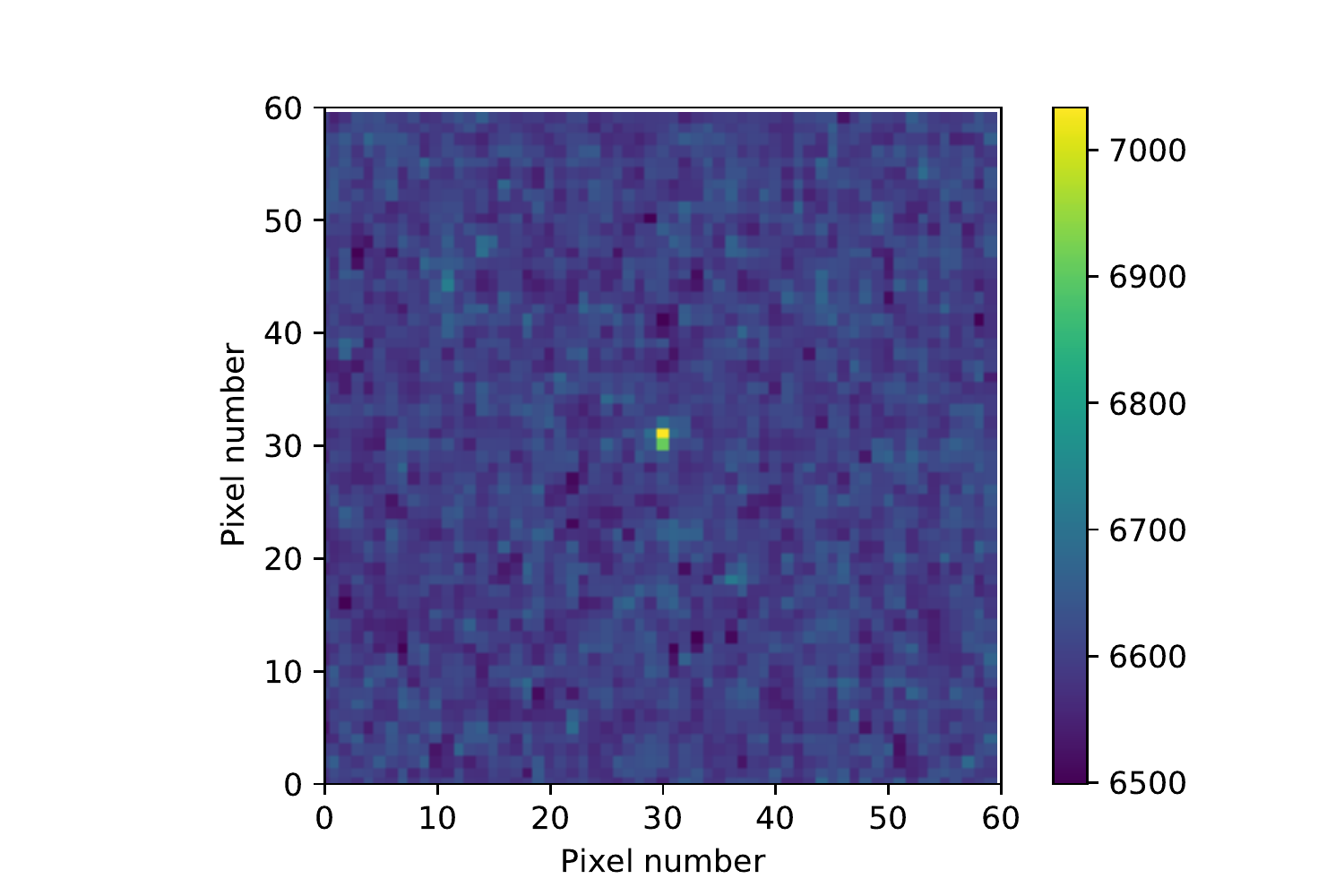}
\caption{The difference image (top) and the original image (bottom) at the location of the candidate transient found in frame \#106, as featured in Figure \ref{diff-sig}.}
\label{zoom}
\end{center}
\end{figure}

Figure \ref{zoom} shows a zoomed in view of one of the five detections, in both the difference image and in the original image.  The difference image shows the signature expected from a transient, with a positive/negative pixel pair in an east-west direction.

Table \ref{detections} lists the properties of the five candidate signals, including positional and photometric information derived from the methods described earlier in the paper.  In Table \ref{detections}, the V magnitude for the candidate is listed for the native 21 ms duration, as well as the V magnitude that would apply assuming that the signal has an intrinsic duration of 1 mas (corresponding to an FRB timescale) and is diluted by the 21 ms duration.

A search for cataloged objects within the error regions listed in Table \ref{detections} was performed using the NASA Extragalactic Database (NED\footnote{https://ned.ipac.caltech.edu/}) and the SIMBAD Astronomical Database\footnote{http://simbad.u-strasbg.fr/simbad/}. 

For the candidate in frame \#97, two objects were found in the error region, WISEA J023331.77$+$013415.8 and GALEXMSC J023331.74$+$013406.5.  As these are only $\sim$10'' apart, these IR and UV objects may be the same object.  For the candidate in frame \#99, one object was found in the error region, WISEA J025140.10$+$025039.5.  For the candidate in frame \#102, one object was found in the error region, WISEA J025546.08$+$025624.3.  For the candidate in frame \#103, no objects were found in the error region.  For the candidate in frame \#106, one object was found in the error region, WISEA J025926.28$+$014656.2.

On the basis that previous high time resolution searches for optical transients have detected objects in Earth orbit, glinting and producing short duration signals, a search of the full catalog of orbiting objects was performed.  The full catalog of Two Line Elements (TLEs) was downloaded from http://www.space-track.org current as of the date of observation.  The Python module Pyephem was utilised to determine if any cataloged objects were in close angular proximity to the candidate events, as seen from the observing location and in the appropriate timerange, searching $\pm$10 seconds around the exposure start time listed in Table \ref{detections}.

For the candidates in frames \#97, \#99, \#103, and \#106, no cataloged objects were within 1$^{\circ}$ near the time of the candidate event.  For the candidate in frame \#102, one cataloged object was within 1$^{\circ}$ near the time of the candidate event, EUTE 25C (cataloged \#27554).

\begin{table*}[h!]
\caption{List of candidate transient signals and their properties}
\centering
\begin{tabular}{|c|c|c|c|c|c|c|c|} \hline \hline
Frame \#&Start           &Pixel      &Peak               &V mag.         &V mag.           &RA (J2000)  & Dec. (2000) \\
        &time (UT)       &coordinates&pixel values       &(21 ms)         &(1 ms)          &            & \\ \hline
97      & 15:06:08$\pm$1 &3691,2635  &575$\pm$29         &6.14$\pm$0.05   &2.83$\pm$0.05    & 02:33:32.7$\pm$23.6 & $+$01:34:11$\pm$4 \\ \hline
99      & 15:16:10$\pm$1 &1185,1294  &1673$\pm$42        &4.98$\pm$0.03   &1.67$\pm$0.03    & 02:51:39.1$\pm$23.6 & $+$02:50:46$\pm$4 \\ \hline
102     & 15:16:32$\pm$1 &1033,2445  &1001$\pm$36        &5.54$\pm$0.04   &2.23$\pm$0.04    & 02:55:44.6$\pm$23.6 & $+$02:56:25$\pm$4 \\ \hline
103     & 15:16:39$\pm$1 &3265,2296  &361$\pm$29         &6.65$\pm$0.09   &3.34$\pm$0.09    & 02:56:06.4$\pm$23.6 & $+$01:50:29$\pm$4 \\ \hline
106     & 15:17:00$\pm$1 &3415,3063  &716$\pm$31         &5.90$\pm$0.04   &2.59$\pm$0.04    & 02:59:28.1$\pm$23.6 & $+$01:46:56$\pm$4 \\ \hline \hline
\end{tabular}
\label{detections}
\end{table*}

\begin{figure*}[h!]
\centering
\includegraphics[width=.4\linewidth]{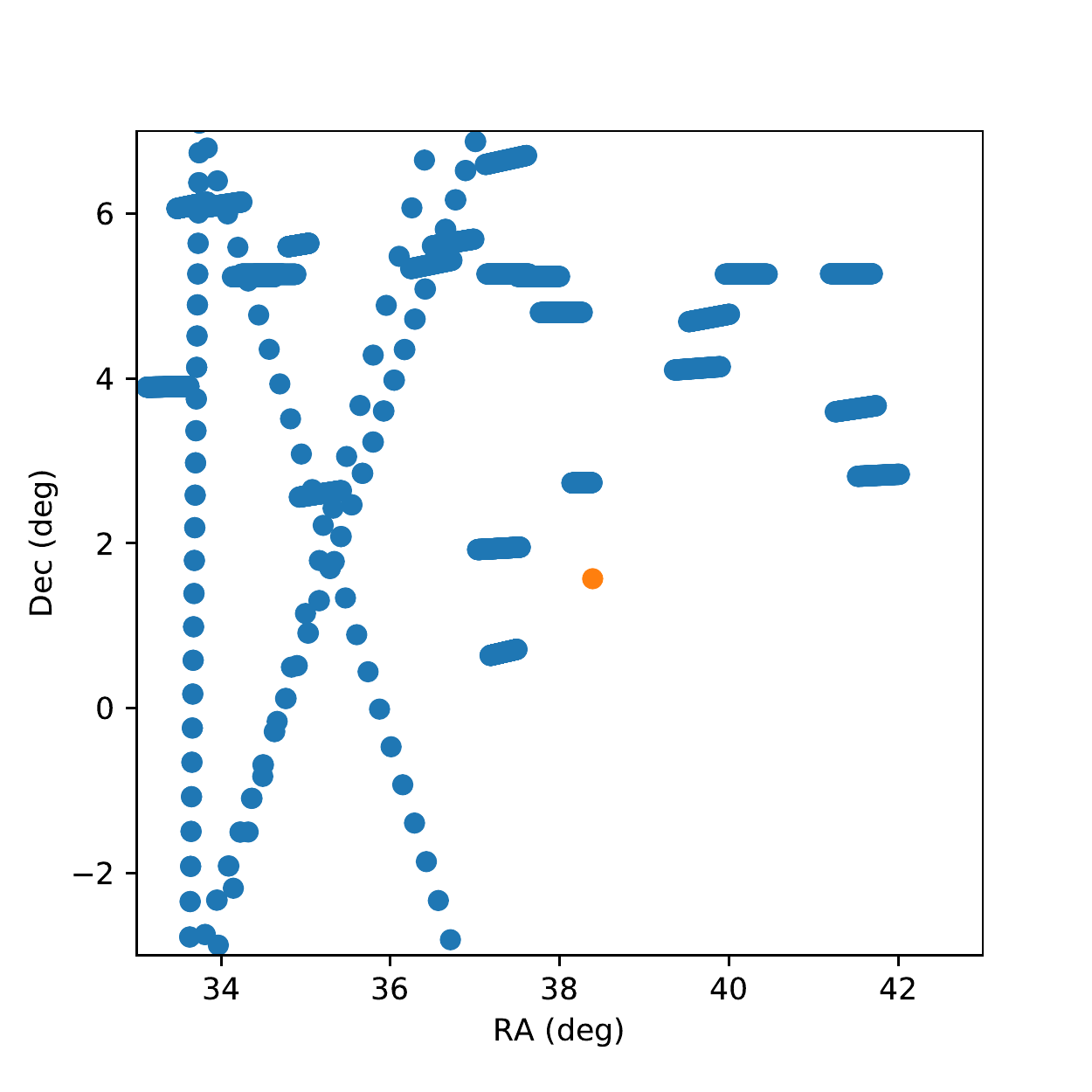}
\label{fig:sub1}
\includegraphics[width=.4\linewidth]{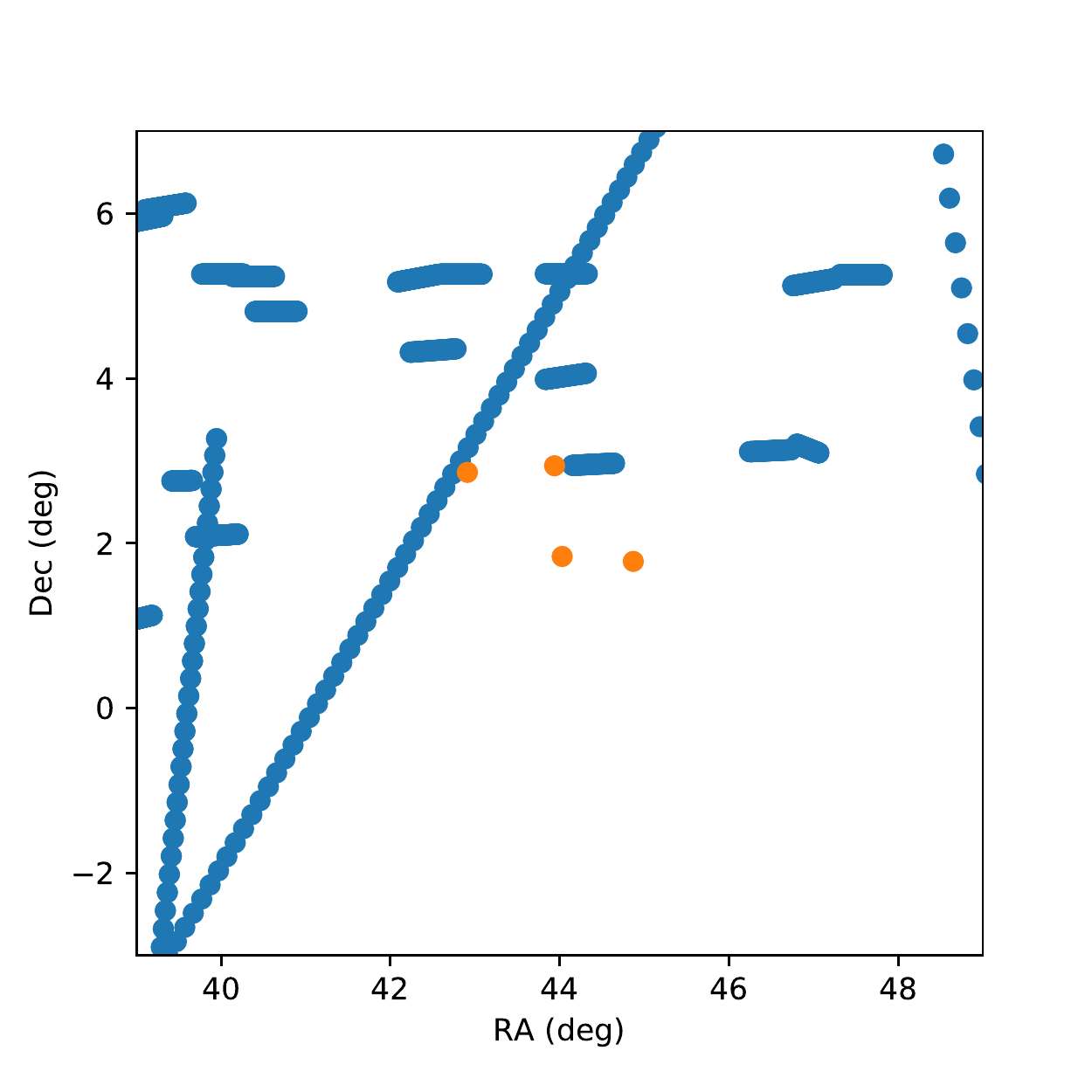}
\label{fig:sub2}
\caption{Tracks of catalogued satellites in the broad vicinity of the candidate detections, as described in the text.  Left, for the detection in frame \#97 as listed in Table \ref{detections}, for the period 15:05:00 - 15:07:00.  Right, for the remaining detections in Table \ref{detections}, for the period 15:15:30 - 15:17:30 and TLEs current as of the observation date.}
\label{satsatday}
\end{figure*}

However, a slightly wider search of the cataloged database shows a number of satellites in the broader vicinity at the times of the observations.  Figure \ref{satsatday} shows the tracks of satellites passing near the candidate detection positions and times, evaluated at one second intervals, utilising the TLEs current at the date of observation.  The observing field and the candidates lie just to the south of the belt of satellites in geosynchronous orbits (centred at $\sim+$5$^{\circ}$ from the observing location).

\begin{figure*}[h!]
\centering
\includegraphics[width=.4\linewidth]{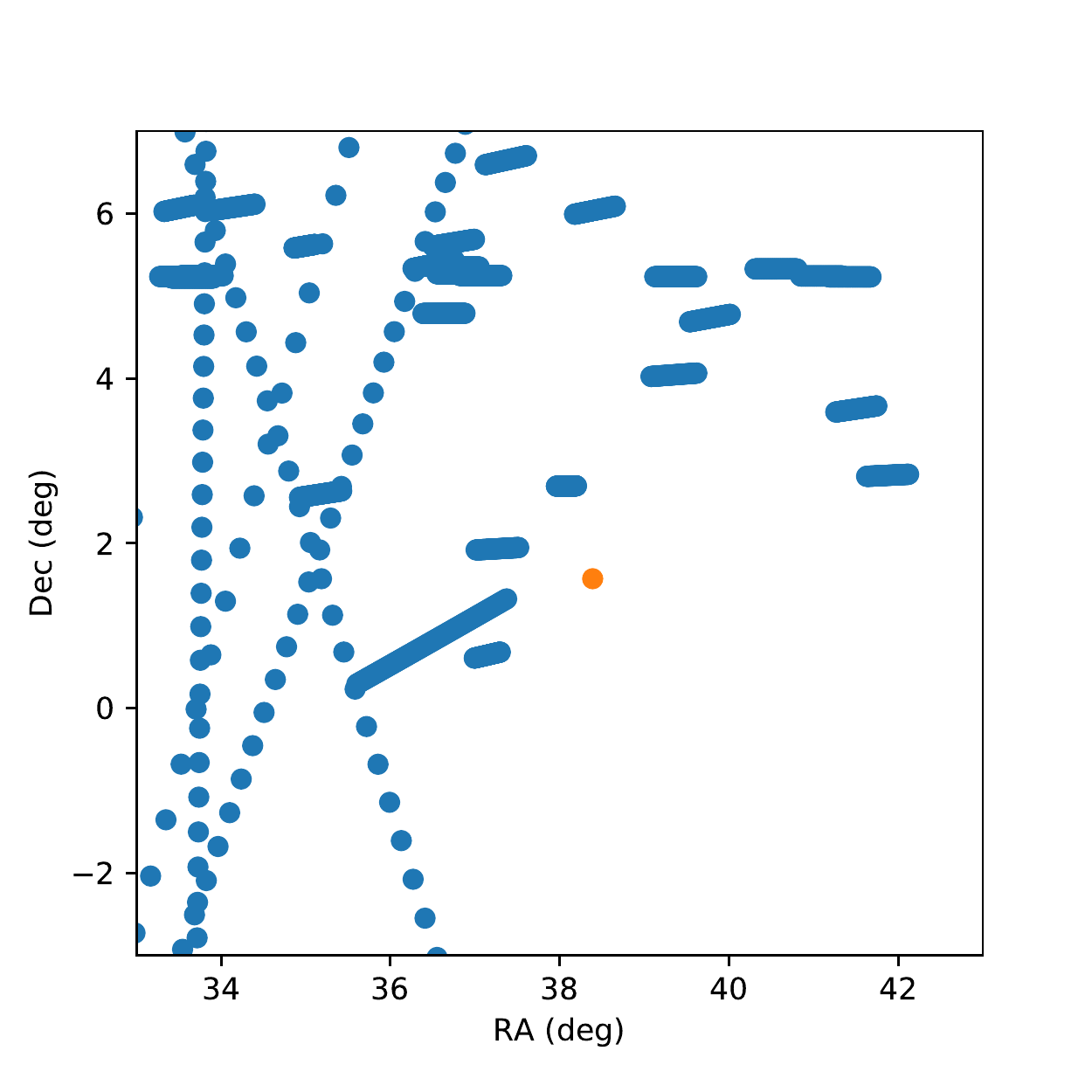}
\label{fig:sub1}
\includegraphics[width=.4\linewidth]{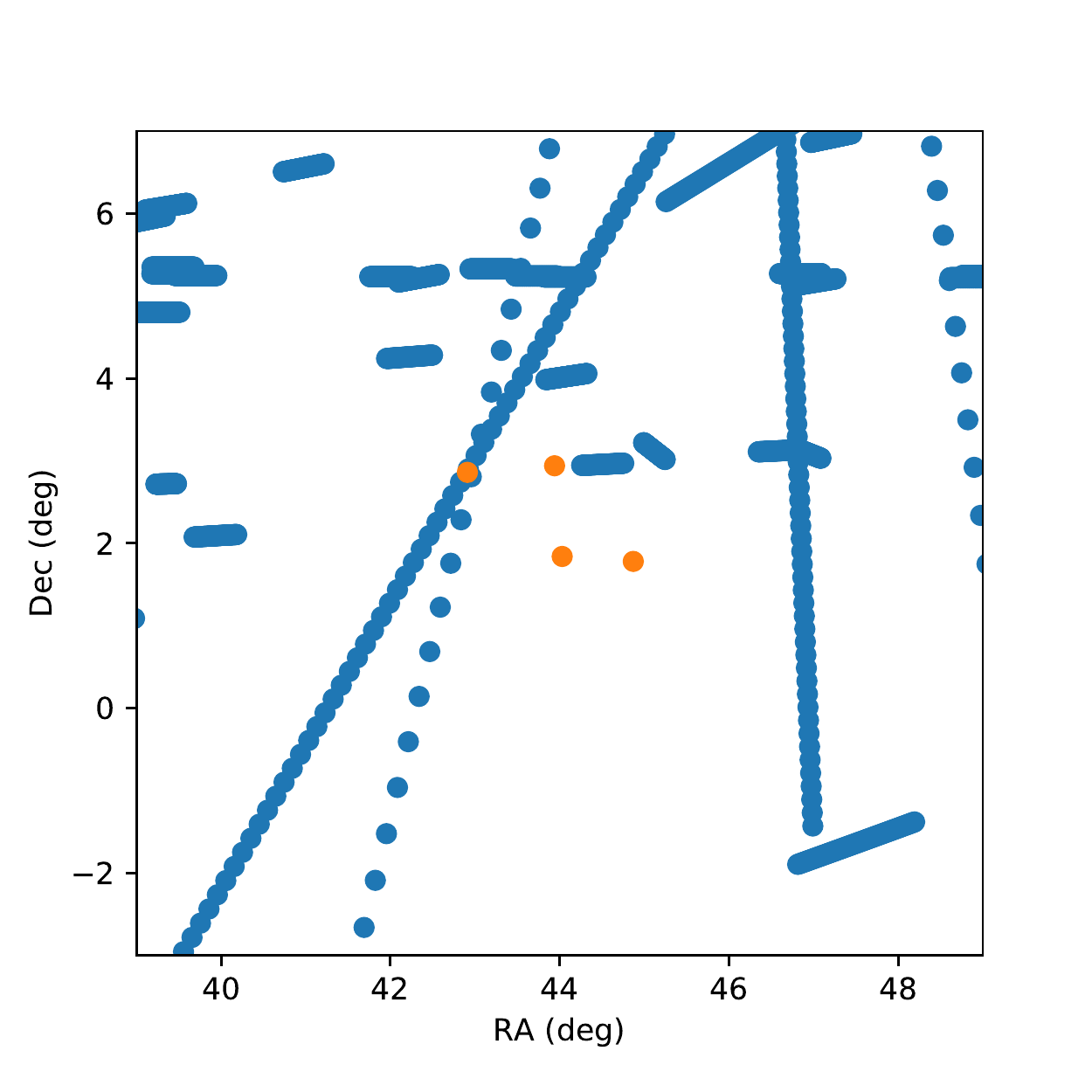}
\label{fig:sub2}
\caption{As for Figure \ref{satsatday}, but with TLEs current for 2019 December 19.}
\label{satsold}
\end{figure*}

Any given update of the TLE catalog does not contain updates on every individual object.  Thus, it is likely that many of the objects seen in the vicinity of the candidates do not have accurate TLEs in the catalog at the time of observation.  To illustrate this, Figure \ref{satsold} shows the same as for Figure \ref{satsatday}, but using the TLE catalog update of 2019 December 19.  Changes to the trajectories of many of the objects are apparent.

Given the uncertainties on the cataloged TLEs, the tracks may coincide more closely with the candidates than indicated in Figure \ref{satsatday}.

In order to verify that the technique of comparing satellites and candidates is valid, a test observation was made with the same setup, but using a 75 mm lens to achieve a wider field, to capture a pass of a known bright (and therefore large) satellite at a known time.  The TLEs of such objects are more frequently updated and are generally more accurate.  A pass of CZ-3 (catalog \#20559) above the horizon at the observing site between approximately 13:42 UT and 14:12 UT was chosen for observation.  CZ-3 passed through the constellation of Canis Major near 13:37 UT and a 5 second exposure was obtained starting at 13:37:42 UT.  The astrometry for the resulting image was solved as described above.  The positions of satellites within the field of view were also determined, using TLEs current at the observation date, also as described above.

\begin{figure*}[h!]
\begin{center}
\includegraphics[width=\textwidth]{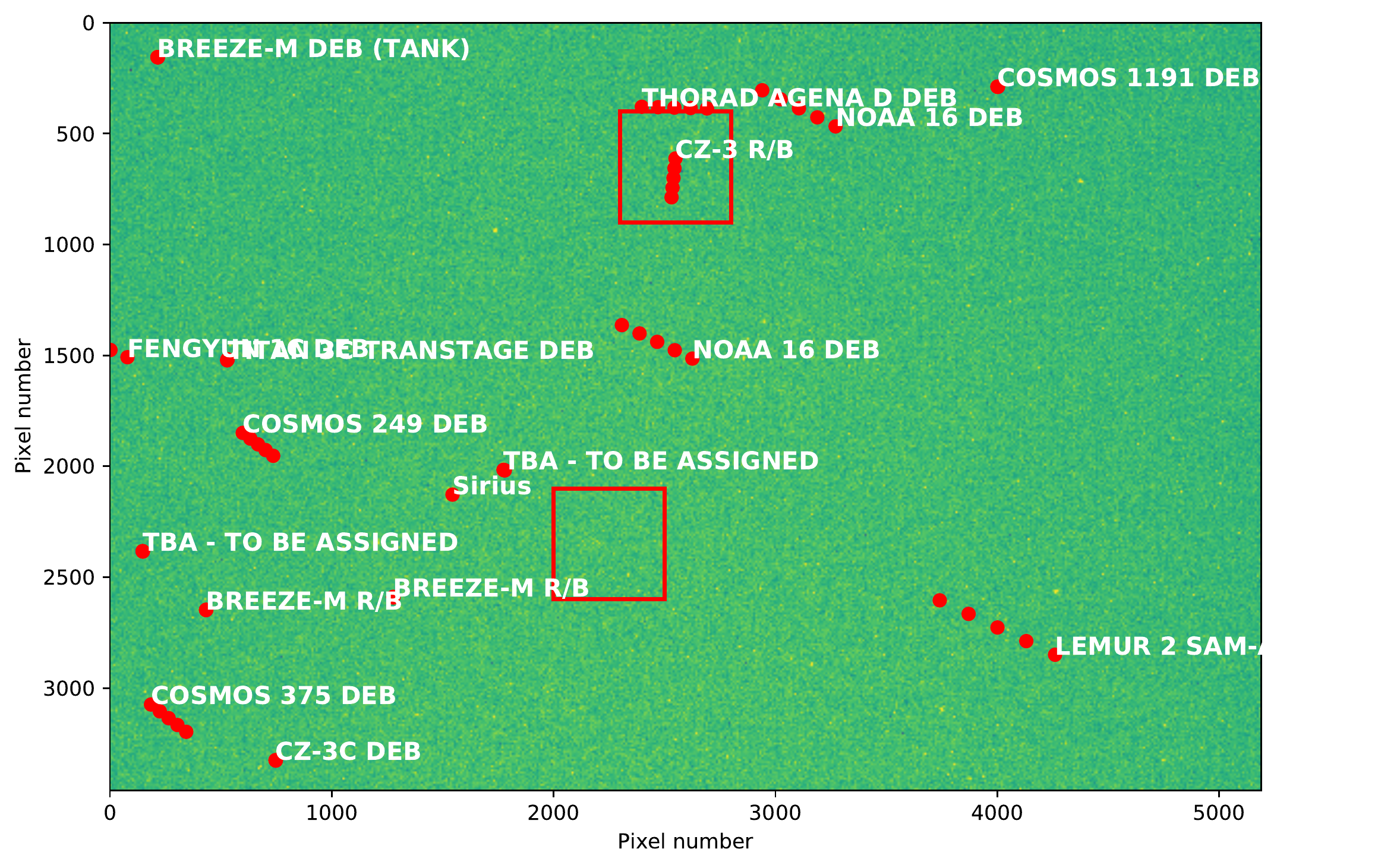}
\caption{A 5 second observation of the constellation of Canis Major during a pass of CZ-3.  The predicted position of Sirius is shown as an astronomical check of the WCS.  Also shown at 1 second intervals over the 5 second period are the positions of the objects contained in the TLE catalog.  Featured near top of image is CZ-3, where the observations and the predicted positions agree.  Also featured near bottom of image is a second satellite, unidentified from the TLEs but suspected to be Cosmos 249; this object is difficult to render for high visibility in the figure but is seen in successive image frames.  It is therefore not a meteor.}
\label{check}
\end{center}
\end{figure*}

Figure \ref{check} shows the image.  The pixel position of Sirius (determined from its RA and Dec and the WCS derived for the image) is shown as a check, which aligns with Sirius on the image.  The observed trajectory of CZ-3 is replicated by its predicted trajectory to within 1 second in time and within several arcminutes in position, indicating that the TLE for CZ-3 at the observation time is an accurate description of its orbit and that the techniques used here to utilise astrometry and TLE information are correct.

Coincidentally, within the same image, a second satellite was detected, highlighted in the second box in Figure \ref{check}, in a polar orbit.  This satellite does not have an obvious counterpart in the TLE catalog.  Of the objects in the field at the time (according to the TLE catalog), the most likely to correspond to the unidentified satellite is Cosmos 249, which has the right sense of direction, moves on a parallel trajectory to the detected object, and moves at the correct angular speed.  If Cosmos 249 is the correct attribution, the TLE error results in $\sim$20 seconds in time and several degrees in position, consistent with the behaviour seen in Figures \ref{satsatday} and \ref{satsold}.

\begin{figure*}[h!]
\begin{center}
\includegraphics[width=\textwidth]{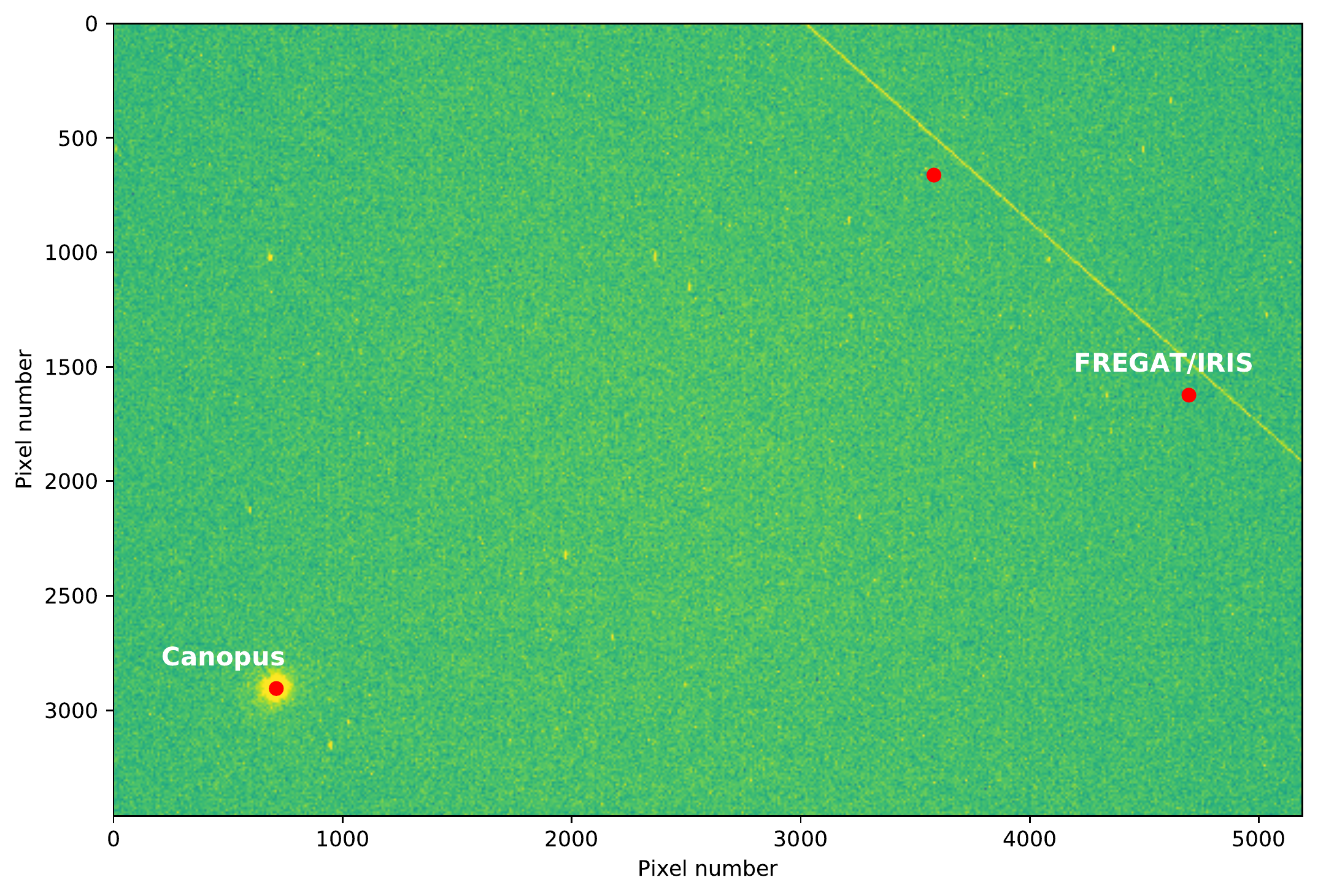}
\caption{A 5 second observation of a pass of FREGAT/IRIS.  The predicted position of Canopus is shown as an astronomical check of the WCS.  Also shown at 1 second intervals over the 5 second period are the positions of FREGAT/IRIS (the only cataloged object in the field at the time).}
\label{satcheck2}
\end{center}
\end{figure*}

In addition, an observation of another well-known object was made with the same setup as for the science observations, with the 500 mm lens, of FREGAT/IRIS (catalog \#35867, another large and bright satellite) using a 5 second exposure starting at 12:19:32 UT, as the object passed the bright star Canopus.  The image is shown in Figure \ref{satcheck2}, again showing that the cataloged TLE accurately predicts the observation, this time using the exact observational setup for the science observations.  The time offset cannot be estimated in this case (as the satellite trail extends beyond the field of view within the 5 seconds), but is at most seconds, and the positional offset is approximately 2 arcminutes.

Finally, a significant source of transient-like signals at the sensor is due to cosmic ray hits on pixels.  Muons at the surface of the Earth caused by cosmic ray air showers occur at a rate of approximately 1${\rm cm}^{-2}$ per minute, so with the sensor area of approximately 3.3${\rm cm}^{2}$, five Muons in 55 seconds of exposure time is a reasonable explanation for the candidates detected here.  A test was conducted by blocking light from the sensor and acquiring a set of 10$\times$30 s dark frames.  These were corrected for bad pixels, as described for the main observations and muon hits were counted and found to be reasonable for five minutes of observation time.  The simultaneous use of a second camera would immediately identify these events and allow them to be rejected.  This is likely to be an essential component of a final system, given the high event rate of cosmic ray particles compared to any expected astrophysical events.

While unlikely to be a contaminant for this experiment, given their slow angular motions on the sky and persistent nature, other Solar System objects, such as asteroids, may be present in the data.

\section{CONCLUSIONS AND FUTURE WORK}
\label{sec:conclusion}
The work described here explores a technique to produce efficient high cadence imaging of short timescale transients, motivated by a desire to detect FRBs at optical wavelengths.  An approach that focuses on a simple COTS pre-prototype system for the generation of test datasets, and an exploration of the calibration and detection methodology, is taken here as a first step toward a more capable system.

The technique has a number of advantages, namely: 1) high cadence imaging via long duration images obtained while the sky drifts across the sensor greatly reduces the data volume (by a factor of 200 for 25 ms cadence from 5 s images); 2) forming the difference images produces a convenient detection space, that imposes a clear signature for transient signals and allows them to be distinguished from other types of signal (satellites and meteors); 3) the technique has relaxed requirements and can scale to a range of telescopes and detectors, obtaining large amounts of time on sky; and 4) a significant system can be produced in a very cost effective manner.

The technique does have drawbacks, however.  The primary limitation is the highly degraded ability to localise any transient signal in the direction of sky drift (right ascension in the case presented here).

Five candidate detections were found over a set of 11 images of 5 s duration.  The analysis presented shows that the candidates are not instrumental in nature.  However, no obvious astronomical counterparts (galactic or extragalactic) are evident within the candidate error regions, although a handful of galaxies are consistent with the error regions for the candidates.  As noted in the introduction, an FRB is predicted to occur every 30 hours in a given 10 sq. deg. area of the sky.  With a field of view of approximately 4.4 sq. deg., the observations described here would expect to contain an FRB approximately every 70 hours.  Thus, the rate of candidates in the data presented here is far higher than any reasonable expectation from our knowledge of FRBs.

Most likely the candidates are muons from cosmic ray showers, or possibly glints from satellites in Earth orbit, given uncertainties on satellite positions.  An analysis of the positions of cataloged satellites relative to the candidate times and positions do not show any close coincidences, however the fact that many cataloged TLEs are generally out of date means that the satellites in the vicinity can be plausibly related to the candidates.  Other high time resolution imaging experiments have commonly detected satellites \citep{2019arXiv191011343R}.  As Earth orbits continue to be populated by objects with poorly understood (infrequently updated) trajectories, the study of optical transients is likely to be impacted.

However, the techniques demonstrated show that if optical bursts are associated with FRBs, then they can be plausibly detected using these techniques.  For example, the faintest detected candidate has a V magnitude of 6.6 in a 21 ms period.  According to the theoretical models of \citet{yang}, this gives an optical flux density limit of 200 Jy for an optical counterpart to a 1 ms FRB.  This is a factor of ten better than was achieved by \citet{Tingay_2019}, for FRB181228.  \citet{yang} predicted an optical flux density of approximately 1 Jy for FRB181228.  Following the results of \citet{Tingay_2019}, for a 200 Jy optical flux density limit, an FRB such as FRB181228 could possibly be detected to a distance of approximately 100 Mpc ($\sim$3 times more distant than the nearest estimated FRB, FRB171020).  Thus, the techniques described here provide a useful capability to explore short duration optical transients at a very modest cost and technical complexity.  Most of the technical complexity revolves around verifying and excluding reflections from satellites and mitigating the effects of a high rate of cosmic ray hits on pixels (both addressed in the future work section, below).

Additionally, given the pre-prototype system described here is constructed from less than \$1000 of commodity equipment and commonly available software and programming environments (including the use of web-based services), the pre-prototype system could form a very capable educational tool at an undergraduate or advanced high school level.  Students would need to grapple with the basics of coordinate systems, astronomical imaging, calibration (photometry and astrometry), signal detection and image processing, and orbital mechanics (to detect and characterise satellites).

Based on the results of this work, several improvements suggest themselves for future work.

\subsection{Future work: pre-prototype}
Given the promising results from even the simple pre-prototype system, some continued testing of the pre-prototype should proceed.  The pre-prototype will be deployed to a dark sky location and run over the course of multiple nights, in order to collect and process a much larger set of images, to more deeply explore the calibration and detection techniques described in this paper.

Given the challenges in verifying and identifying astrophysically interesting events from the candidates, additional verification steps will be required.  An obvious addition to the pre-prototype would be a second camera, pointed to the same part of the sky and run simultaneously.  Ideally, some geographic distance between the camera would allow near-field objects such as satellites to be identified.  Any real astrophysical transients would appear at the same RA and Dec and at the same time.  Filtering cosmic ray hits will greatly reduce the candidate event rate that requires detailed consideration.

The techniques described here ultimately aim to produce candidate transient-host associations.  If the improved pre-prototype system, using two cameras, greatly reduces the number of candidate detections, the confidence of these candidate detections will be greatly increased.  In this case, with a very small number of candidate detections to follow up in detail, candidate associations would be considered to be any object within the error region.  Further targeted high time resolution radio and optical observations would be conducted to possibly detect repeats and also to better localise the host to secure transient-host associations.

A clear next step for the pre-prototype (and beyond), when undertaking long series of exposures to obtain significant time on sky, is to develop a more automated data processing pipeline involving many fewer manual steps.  For example, while www.astrometry.net provides a convenient service for testing, or for amateur/student projects, a better automated approach at scale may be to use the Astromatic/SCAMP package \citep{2006ASPC..351..112B}.

\subsection{Future work: prototype}
The planned next step is to scale up to a prototype system.  The success of the COTS approach suggests that it represents a reasonable path to follow to prototype.

With this in mind, a concept for a more capable prototype system has been developed.  This system is based around a Celestron 279 mm aperture, f/2.2 Rowe-Ackermann Astrograph (620 mm focal length).  This telescope is optimised for a flat response over a wide field of view and large format commercially available sensors.  For example, coupled with a 36 mm $\times$ 24 mm sensor, the imaged field of view is approximately 3.3$^{\circ}\times$2.2$^{\circ}$ (7.3 sq. deg.).  With a typical pixel size, the 620 mm focal length results in a sky drift rate across the sensor at the equator of 75 ms/pixel, thus a transient duration sensitivity of 75 ms.  Active driving at the same rate as the pre-prototype system would result in a transient duration sensitivity of $<$20 ms.

With a factor of 3.5 improvement in the aperture diameter compared to the pre-prototype, an approximate order of magnitude improvement in sensitivity would be expected, or approximately three magnitudes of improvement in V.

\subsection{Future work: SkyMapper}
In order to significantly improve the sensitivity of the prototype system, large-scale professional facilities are required.  The general techniques explored here can be used at such facilities, with the same benefits.

With this in mind, an exploration of the feasibility of a drift imaging experiment using the SkyMapper telescope \citep{2007PASA...24....1K}, based at Siding Spring Observatory, has been undertaken with the SkyMapper team.  With a 1.35 m diameter mirror at f/4.8 (F=6.5 m) and a 0.5''/pixel sensor scale over a 2.4$^{\circ}\times$2.4$^{\circ}$ field of view, the sky drift rate at the equator is approximately 12 ms/pixel.

With a factor of 4.8 improvement in the aperture diameter compared to the proposed prototype, an approximate factor of 20 improvement in sensitivity can be achieved, or approximately 3.5 magnitudes of improvement in V (improvement of 6.5 magnitudes of improvement over the pre-prototype results).

Three nights of dark time with SkyMapper will be obtained to run this experiment.

\begin{acknowledgements}
I thank the anonymous referee for comments that improved this paper.  This research has made use of the SIMBAD database, operated at CDS, Strasbourg, France. This research has made use of the NASA/IPAC Extragalactic Database (NED), which is funded by the National Aeronautics and Space Administration and operated by the California Institute of Technology.  I gratefully acknowledge the donation of the ``herritage'' mount that was refurbished and used in this work, by Mr Roy Deering from the Astronomical Group of Western Australia (AGWA).
\end{acknowledgements}

\bibliographystyle{pasa-mnras}
\bibliography{custom}

\begin{thebibliography}{}
\makeatletter
\relax
\def\mn@urlcharsother{\let\do\@makeother \do\$\do\&\do\#\do\^\do\_\do\%\do\~}
\definecolor{darkblue}{rgb}{0,0,0.597656}
\def\mndoi{\begingroup\mn@urlcharsother \@ifnextchar [ {\mndoi@} {\mndoi@[]}}
\def\mndoi@[#1]#2{\def\@tempa{#1}\ifx\@tempa\@empty \href
  {http://dx.doi.org/#2} {\textcolor{darkblue}{doi:#2}}\else \href
  {http://dx.doi.org/#2} {\textcolor{darkblue}{#1}}\fi \endgroup}
\def\mn@eprint#1#2{\mn@eprint@#1:#2::\@nil}
\def\mn@eprint@arXiv#1{\href {http://arxiv.org/abs/#1} {{\tt arXiv:#1}}}
\def\mn@eprint@dblp#1{\href {http://dblp.uni-trier.de/rec/bibtex/#1.xml}
  {dblp:#1}}
\def\mn@eprint@#1:#2:#3:#4\@nil{\def\@tempa {#1}\def\@tempb {#2}\def\@tempc
  {#3}\ifx \@tempc \@empty \let \@tempc \@tempb \let \@tempb \@tempa \fi \ifx
  \@tempb \@empty \def\@tempb {arXiv}\fi \@ifundefined
  {mn@eprint@\@tempb}{\@tempb:\@tempc}{\expandafter \expandafter \csname
  mn@eprint@\@tempb\endcsname \expandafter{\@tempc}}}

\bibitem[\protect\citeauthoryear{{Andreoni} et~al.,}{{Andreoni}
  et~al.}{2020}]{2020MNRAS.491.5852A}
{Andreoni} I.,  et~al., 2020, \mndoi [\mnras] {10.1093/mnras/stz3381}, \href
  {https://ui.adsabs.harvard.edu/abs/2020MNRAS.491.5852A} {491, 5852}

\bibitem[\protect\citeauthoryear{{Bandura} et~al.,}{{Bandura}
  et~al.}{2014}]{2014SPIE.9145E..22B}
{Bandura} K.,  et~al., 2014, {Canadian Hydrogen Intensity Mapping Experiment
  (CHIME) pathfinder}.
p. 914522, \mndoi{10.1117/12.2054950}

\bibitem[\protect\citeauthoryear{{Bertin}}{{Bertin}}{2006}]{2006ASPC..351..112B}
{Bertin} E.,  2006, {Automatic Astrometric and Photometric Calibration with
  SCAMP}.
p.~112

\bibitem[\protect\citeauthoryear{{Burke-Spolaor}}{{Burke-Spolaor}}{2018}]{2018NatAs...2..845B}
{Burke-Spolaor} S.,  2018, \mndoi [Nature Astronomy]
  {10.1038/s41550-018-0630-x}, \href
  {https://ui.adsabs.harvard.edu/abs/2018NatAs...2..845B} {2, 845}

\bibitem[\protect\citeauthoryear{{Chawla} et~al.,}{{Chawla}
  et~al.}{2017}]{2017ApJ...844..140C}
{Chawla} P.,  et~al., 2017, \mndoi [\apj] {10.3847/1538-4357/aa7d57}, \href
  {https://ui.adsabs.harvard.edu/abs/2017ApJ...844..140C} {844, 140}

\bibitem[\protect\citeauthoryear{{Groot}, {Bloemen}  \& {Jonker}}{{Groot}
  et~al.}{2019}]{2019lsof.confE..33G}
{Groot} P.,  {Bloemen} S.,   {Jonker} P.,  2019, in The La Silla Observatory -
  From the Inauguration to the Future. p.~33, \mndoi{10.5281/zenodo.3471366}

\bibitem[\protect\citeauthoryear{{Hardy} et~al.,}{{Hardy}
  et~al.}{2017}]{2017MNRAS.472.2800H}
{Hardy} L.~K.,  et~al., 2017, \mndoi [\mnras] {10.1093/mnras/stx2153}, \href
  {https://ui.adsabs.harvard.edu/abs/2017MNRAS.472.2800H} {472, 2800}

\bibitem[\protect\citeauthoryear{{H{\o}g} et~al.,}{{H{\o}g}
  et~al.}{2000}]{2000A&A...355L..27H}
{H{\o}g} E.,  et~al., 2000, \aap, \href
  {https://ui.adsabs.harvard.edu/abs/2000A&A...355L..27H} {355, L27}

\bibitem[\protect\citeauthoryear{{Karpov} et~al.,}{{Karpov}
  et~al.}{2017}]{2017ASPC..510..526K}
{Karpov} S.,  et~al., 2017, {Mini-MegaTORTORA Wide-Field Monitoring System with
  Subsecond Temporal Resolution: Observation of Transient Events}.
p.~526

\bibitem[\protect\citeauthoryear{{Keller} et~al.,}{{Keller}
  et~al.}{2007}]{2007PASA...24....1K}
{Keller} S.~C.,  et~al., 2007, \mndoi [\pasa] {10.1071/AS07001}, \href
  {https://ui.adsabs.harvard.edu/abs/2007PASA...24....1K} {24, 1}

\bibitem[\protect\citeauthoryear{{Lang}, {Hogg}, {Mierle}, {Blanton}  \&
  {Roweis}}{{Lang} et~al.}{2010}]{2010AJ....139.1782L}
{Lang} D.,  {Hogg} D.~W.,  {Mierle} K.,  {Blanton} M.,   {Roweis} S.,  2010,
  \mndoi [\aj] {10.1088/0004-6256/139/5/1782}, \href
  {https://ui.adsabs.harvard.edu/abs/2010AJ....139.1782L} {139, 1782}

\bibitem[\protect\citeauthoryear{{Marcote} et~al.,}{{Marcote}
  et~al.}{2020}]{2020Natur.577..190M}
{Marcote} B.,  et~al., 2020, \mndoi [\nat] {10.1038/s41586-019-1866-z}, \href
  {https://ui.adsabs.harvard.edu/abs/2020Natur.577..190M} {577, 190}

\bibitem[\protect\citeauthoryear{{Paterson}}{{Paterson}}{2019}]{2019IAUS..339..203P}
{Paterson} K.,  2019, in {Griffin} R.~E.,  ed.,  IAU Symposium Vol. 339,
  Southern Horizons in Time-Domain Astronomy. pp 203--203,
  \mndoi{10.1017/S1743921318002594}

\bibitem[\protect\citeauthoryear{{Petroff} et~al.,}{{Petroff}
  et~al.}{2016}]{2016PASA...33...45P}
{Petroff} E.,  et~al., 2016, \mndoi [\pasa] {10.1017/pasa.2016.35}, \href
  {https://ui.adsabs.harvard.edu/abs/2016PASA...33...45P} {33, e045}

\bibitem[\protect\citeauthoryear{{Richmond} et~al.,}{{Richmond}
  et~al.}{2019}]{2019arXiv191011343R}
{Richmond} M.~W.,  et~al., 2019, arXiv e-prints, \href
  {https://ui.adsabs.harvard.edu/abs/2019arXiv191011343R} {p. arXiv:1910.11343}

\bibitem[\protect\citeauthoryear{{Shearer} \& {Connor}}{{Shearer} \&
  {Connor}}{2018}]{2018IAUS..337..191S}
{Shearer} A.,  {Connor} E. O.~.,  2018, in {Weltevrede} P.,  {Perera} B.~B.~P.,
   {Preston} L.~L.,   {Sanidas} S.,  eds,  IAU Symposium Vol. 337, Pulsar
  Astrophysics the Next Fifty Years. pp 191--194,
  \mndoi{10.1017/S174392131700998X}

\bibitem[\protect\citeauthoryear{{The CHIME/FRB Collaboration} et~al.,}{{The
  CHIME/FRB Collaboration} et~al.}{2020}]{2020arXiv200110275T}
{The CHIME/FRB Collaboration} et~al., 2020, arXiv e-prints, \href
  {https://ui.adsabs.harvard.edu/abs/2020arXiv200110275T} {p. arXiv:2001.10275}

\bibitem[\protect\citeauthoryear{Tingay \& Yang}{Tingay \&
  Yang}{2019}]{Tingay_2019}
Tingay S.~J.,  Yang Y.-P.,  2019, \mndoi [The Astrophysical Journal]
  {10.3847/1538-4357/ab2c6e}, 881, 30

\bibitem[\protect\citeauthoryear{{Wenger} et~al.,}{{Wenger}
  et~al.}{2000}]{2000A&AS..143....9W}
{Wenger} M.,  et~al., 2000, \mndoi [\aaps] {10.1051/aas:2000332}, \href
  {https://ui.adsabs.harvard.edu/abs/2000A&AS..143....9W} {143, 9}

\bibitem[\protect\citeauthoryear{{Yang}, {Zhang}  \& {Wei}}{{Yang}
  et~al.}{2019}]{yang}
{Yang} Y.-P.,  {Zhang} B.,   {Wei} J.-Y.,  2019, \mndoi [\apj]
  {10.3847/1538-4357/ab1fe2}, \href
  {https://ui.adsabs.harvard.edu/abs/2019ApJ...878...89Y} {878, 89}

\makeatother
\end{thebibliography}

\end{document}